\documentclass[fleqn,usenatbib]{mnras}

\usepackage[T1]{fontenc}
\usepackage{ae,aecompl}

\usepackage{graphicx}	
\usepackage{amsmath}	
\usepackage{amssymb}	

\title[Apparent transients mapping the near-Earth plasma]{Apparent radio transients mapping the near-Earth plasma environment}

\author[M. J. Kuiack et al.]{
Mark J. Kuiack,$^{1}$\thanks{E-mail: m.j.kuiack@uva.nl}
Ralph A.M.J. Wijers,$^{1}$
Aleksandar Shulevski,$^{1}$\newauthor
Antonia Rowlinson,$^{1,2}$
\\
$^{1}$ Anton Pannekoek Institute, University of Amsterdam, Science Park 904, 1098 XH Amsterdam, The Netherlands \\
$^{2}$ ASTRON, The Netherlands Institute for Radio Astronomy, Postbus 2, 7990 AA, Dwingeloo, The Netherlands \\
}

\date{Accepted XXX. Received YYY; in original form ZZZ}

\pubyear{2021}

\begin{document}
\label{firstpage}
\pagerange{\pageref{firstpage}--\pageref{lastpage}}
\maketitle

\begin{abstract}
We report the discovery of bright, fast, radio flashes lasting tens of seconds with the AARTFAAC high-cadence all-sky survey at 60\,MHz. The vast majority of these coincide with known, bright radio sources that brighten by factors of up to 100 during such an event. We attribute them to magnification events induced by plasma near the Earth, most likely in the densest parts of the ionosphere. They can occur both in relative isolation, during otherwise quiescent ionospheric conditions, and in large clusters during more turbulent ionospheric conditions. Using a toy model, we show that the likely origin of
the more extreme (up to a factor of 100 or so) magnification events likely originate
in the region of peak electron density in the ionosphere, at an altitude of 300--400\,km.
Distinguishing these events from genuine astrophysical transients is imperative for future surveys searching for low frequency radio transient at timescales below a minute. 
\end{abstract}

\newpage


\section{Introduction}
\label{sec:intro}

In recent years low-frequency  (<200 MHz) transient surveys have been completed with the Long Wavelength Demonstrator Array (LWDA) \citep{2010AJ....140.1995L}, the Murchinson Widefield Array (MWA) \citep{2014MNRAS.438..352B, 2019MNRAS.482.2484B}, Long Wavelength Array (LWA1) \citep{2015JAI.....450004O}, Low-Frequency Array (LOFAR) \citep{2016MNRAS.459.3161C},  Owens Valley Radio Observatory Long Wavelength Array (OVRO-LWA) \citep{2019ApJ...886..123A}. 
These focus on the search for bright transient or variable emission, with the hope of discovering new extremely energetic processes taking place in the Universe. 
Previously, we have reported the discovery of extremely high fluence giant pulses from PSR\,B0950$+$08 \citep{2020arXiv200300720K} during the course of an ongoing blind transients survey with the Amsterdam-ASTRON Radio Transient Facility And Analysis Center \citep[AARTFAAC;][]{2014A&A...568A..48P,2016JAI.....541008P}. 
We were easily able to  associate this emission to the specific astrophysical source, because the signal repeated and the source properties and behaviour had been well studied. 
However, the goal of blind transient surveys such as AARTFAAC is also to search for as-yet unknown phenomena. 
This necessarily means that any candidate events discovered, may not be so easily associated with known objects, or known processes. 

An example is the discovery  of an unknown transient source at low frequency in the LOFAR North Celestial Pole survey \citep{2016MNRAS.456.2321S}. 
No strong association to a progenitor, host location, or known astrophysical phenomenon could be made, thus the primary validation is to minimize the chance it could be an imaging artefact. 
Once we are convinced that an event is not an instrumental or data processing artefact, we can claim detection of a source.
However, we then need to check whether the event could not be a noise spike associated with normal measurement error distributions, or variability induced by the ionosphere, which has particularly strong effect in the AARTFAAC frequency range (30--80\,MHz), \citep{2018A&A...615A.179D}.
Fortunately, a high cadence transient survey generates a huge number of measurements (one per second per frequency band in the case of AARTFAAC) and thus allows us to characterise the noise distribution and ionospheric behaviour quite well.  
We have already studied the variations due to the quiescent ionosphere and noise quite extensively when constructing the AARTFAAC catalogue \citep{10.1093/mnras/sty2810}, and when studying PSR\,B0950$+$08 \citep{2020arXiv200300720K}. 
The bottom line of our findings was that the measurement noise does not play a dominant role in the sources we detect almost by definition, because we set the detection threshold at 5--8 times the rms. noise; the ionosphere-induced variations in quiescence have an amplitude of 20--30\%, on a timescale of 10--20\,min at 60\,MHz.

However, in analyzing 545 hours for the AARTFAAC transient survey \citep{2020arXiv200313289K}, more exotic ionospheric behaviours were observed, which confound the detection of transient signals intrinsically emitted from astrophysical sources. 
Studying these conditions, and their effects on the statistics of measured light-curves in detail can inform the criteria required for low frequency transient candidate selection. 
For example \cite{2019ApJ...874..151V} used the kurtosis of the flux measurement distribution to argue that an event which was an outlier among a population of scintillating sources it was more likely a astrophysical transient. Strongly  scintillating sources commonly exhibit many scintels over a period of time and therefore the total flux density distribution is  skewed toward higher values. Their claimed transient is a single flash, so the total measurement distribution is not skewed. This transient claim does not seem to consider the possibility of single isolated scintels, our work seeks to explore and and differentiate the statistical properties of singly and multiply magnified scintillating sources.
Further, in attempting to detect transient and variable sources \cite{2019A&C....27..111R} explore the distributions of two parameters, the weighted reduced $\chi^2$ and the variability index, across all sources observed during an observation. Simulations illustrated how distinct transient types would appear as outliers in this distribution. We hope to extend the toolkit of transient detection by introducing multi-frequency pulse shape and delay analysis to the light-curve of each source.
Additionally, studying periods of strong scintillation gives insight into the dynamics of the ionosphere \citep{1996ApOpt..35..986H}, and due to the coupling of the magnetosphere-ionosphere/thermosphere system \citep{2016JGRA..12111861T} the broader near-Earth plasma environment as well \citep{2013Ge&Ae..53..275B, 2019Ge&Ae..59..554D}.

Here we report the discovery of rapid, large-amplitude variations in the 60\,MHz flux densities of normally constant sources on short timescales: sources show apparent flare-ups in brightness of a factor few to over 100 on time scales of a few tens of seconds.
They can occur both in relative isolation, when other nearby sources show their expected quiescent behaviour, or in large groups over a fairly large area of sky, for a period lasting hours. 
In most cases careful study reveals that they are due to exceptional effects of relatively nearby plasma. 
The large regions of strong scintillation are potentially due to, and a new method for observing, a travelling ionospheric disturbance (TID).

In Section \ref{sec:obs} we describe the instrument set-up, the set of observations, and the methodology used to analyse light curves and detect isolated bursts. 
Then in Section \ref{sec:results} we present our results, both a qualitative and quantitative description of light curve features that mimick short-duration transients and variables. 
Next, in Section \ref{sec:discussion} we explore the potential cause. Lastly, in Section \ref{sec:conclusion} we give our conclusions.

\section{Instrument,  Observation, and Methods}
\label{sec:obs}

AARTFAAC is a project whose aim is to survey the Northern Hemisphere for bright (tens of Jy), low-frequency (30--80\,MHz) radio transients, with timescales upwards of one second \citep{2016JAI.....541008P}. 
In the AARTFAAC-6 configuration, data are streamed directly from 288 low band antennas (LBA) in the six central stations of the Low Frequency Array \mbox{\citep[LOFAR;][]{2013A&A...556A...2V}}, the so called ``superterp,'' providing a maximum baseline of 300\,m and resolution of 1$^\circ$ at 60\,MHz. The data are processed in parallel during other ongoing LOFAR-LBA observations, or designated ``filler'' time,  employing an independent correlator, and subsequent calibration and imaging pipelines on a dedicated compute and storage cluster \mbox{\citep{2014A&A...568A..48P}}. 
The data are recorded with a one second time integration, in 16 separate sub-bands, each with a bandwidth of 195.3 kHz, and arranged in two contiguous sequences of eight sub-bands covering 57.6--59.0 MHz, and 61.1--62.5 MHz. 

Real-time, streaming, complex-gain calibration is performed with custom built software (described fully in \citep{2014A&A...568A..48P}), which employs a multi-source self-calibration scheme \citep{2009ITSP...57.3512W}. The sky model consists of five sources Cassiopeia A, Cygnus A, Taurus A, Virgo A, and the Sun. These are each modeled with a single  Gaussian component, an accurate enough approximation at AARTFAAC-6 resolution. At each time-step, both direction independent and direction dependent gain solutions are calculated, however, converging gain solutions are propagated forward in time to reduce the required iterations and thus computation time. Once the visibilities are calibrated, the bright sources used in calibration are subtracted.

The calibrated visibilities are then streamed to the archive disk and our custom built imaging pipeline. The imaging pipeline simply applies flags to visibilities with baselines shorter than $10\lambda$ to reduce the appearance of bright diffuse galactic emission, then performs frequency dependent gridding and Fourier transform. This creates a $1024\times1024$ pixel image for each of the two 1.5\,MHz bands, centered at 58.3 and 61.8 MHz.
Flux scaling is done in the image plane by comparing many flux density measurements from persistent sources in the image, to the AARTFAAC catalogue \citep{10.1093/mnras/sty2810}. 
The image stream can then be saved to disk for off-line transient analysis, and/or streamed to the AARTFAAC website\footnote{\url{www.aartfaac.org/live}, video stream hosted by YouTube} for live data quality inspection and public outreach. 
The total duration of data analyzed in this work was 545.25 hours, collected between 2016-8-31 and 2019-9-14. 

With 8 subbands (1.56\,MHz) and 1 second integration, AARTFAAC-6 is confusion noise limited, the theoretical thermal noise limit is 1\,Jy, while with a resolution of 1$^\circ$ the confusion noise is 10.4\,Jy/beam \citep[Eqn. 6;][]{2013A&A...556A...2V}. 
The typical background rms noise in the images, averaged across the field of view and the entire survey is 7.5\,Jy/beam, yielding typical 5 and 8 $\sigma$  detection thresholds of 37.5 and 60\,Jy.

Light curves are generated from the images by the LOFAR Transient Pipeline v3.1.1\footnote{\url{https://github.com/transientskp/tkp}} \citep[TraP;][and references therein]{2015A&C....11...25S}.
TraP is a Python and SQL based transient detection pipeline for image-based surveys.
The only non-standard parameters used in TraP are the one that sets the detection threshold as a function of the local rms.\ noise level, \texttt{detection\_threshold = $5\sigma$}, and one that forces the use of a fixed-width 2D Gaussian beam during source extraction. This is set by the \texttt{force\_beam = True} parameter in the TraP job configuration file, which locks the widths to the semi-major and semi-minor axis of the restoring beam, read from the fits header.

\subsection{Light curve feature analysis}
\label{sec:analysis}

When analyzing a potential transient event, we are interested in measuring important quantities such as the peak flux in each frequency, the delay between frequencies, and the duration. 
This is done automatically using the \texttt{scipy.signal} \citep{mckinney-proc-scipy-2010} modules \texttt{find\_peaks} and  \texttt{peak\_widths}. The algorithm for detecting burst-like features in light curves is as follows:
\begin{enumerate}
    \item First, the data are regularized in time. By default measurements are only recorded when a signal greater than $5\sigma$ is detected. We therefore fill the light curve between the min and max time stamp with zeros, then make a copy of the light curve, smoothed with a flat 3-second kernel to cover missing time steps, and decrease noise. 
    \item Peaks are detected on the smoothed light curve with \texttt{find\_peaks} using the parameters:
    \begin{itemize}
        \item \texttt{min\_width = 5} [seconds]: The minimum FWHM of light curve features to be considered. Together with the 3 second smoothing, this ensures that single-bin bright noise spikes are not considered. 
        \item \texttt{max\_width = None}: The maximum FWHM of light curve feature to be considered. 
        \item \texttt{min\_prominence = 30} [Jy]: The minimum difference in height between a light curve feature and its surroundings. Correctly detecting isolated bursts requires that the minimum prominence be larger than the typical measurement variability, reducing the detection of noise spikes on top of smooth light curves (false positives),  but small enough that true positives are not ignored. Therefore this is manually tuned to approximately the minimum height of a true positive.
        \item \texttt{max\_prominence = None}: The maximum difference in height between a light curve feature and its surrounding. 
        \item \texttt{rel\_height = 0.9}: The relative height of the feature where the width is measured. This number corresponds to the proportion above, therefore 1.0 is the base, and 0.0 the peak. 
    \end{itemize}
    \item Peaks are associated using the central time, matching nearest neighbours within a delay interval of --120 to +20 seconds  between the 61.8 and 58.3\,MHz light curves (120\,s delay would result from a DM of $900 \mathrm{~pc~cm^{-3}}$). One-to-many matches are eliminated by choosing the pair with the minimum delay.
    \item Highly skewed peaks, where the time difference between the peak and mid-point are greater than 5 seconds, are removed. This ensures that we are selecting only isolated peaks. Consequently, we may also be removing real  transients that follow a fast rise-exponential decay behaviour or series of
    peaks that are overlapping.
    \item  Lastly, once the features are matched between frequencies, the effective duration of the burst is measured on the smoothed light curve at \texttt{rel\_height = 0.9}, as is the delay (the difference between the midpoints in time), and the peak is measured from the original light curve, which consists of measurements over $5\sigma$.
\end{enumerate}

\begin{figure}
\includegraphics[width=\columnwidth]{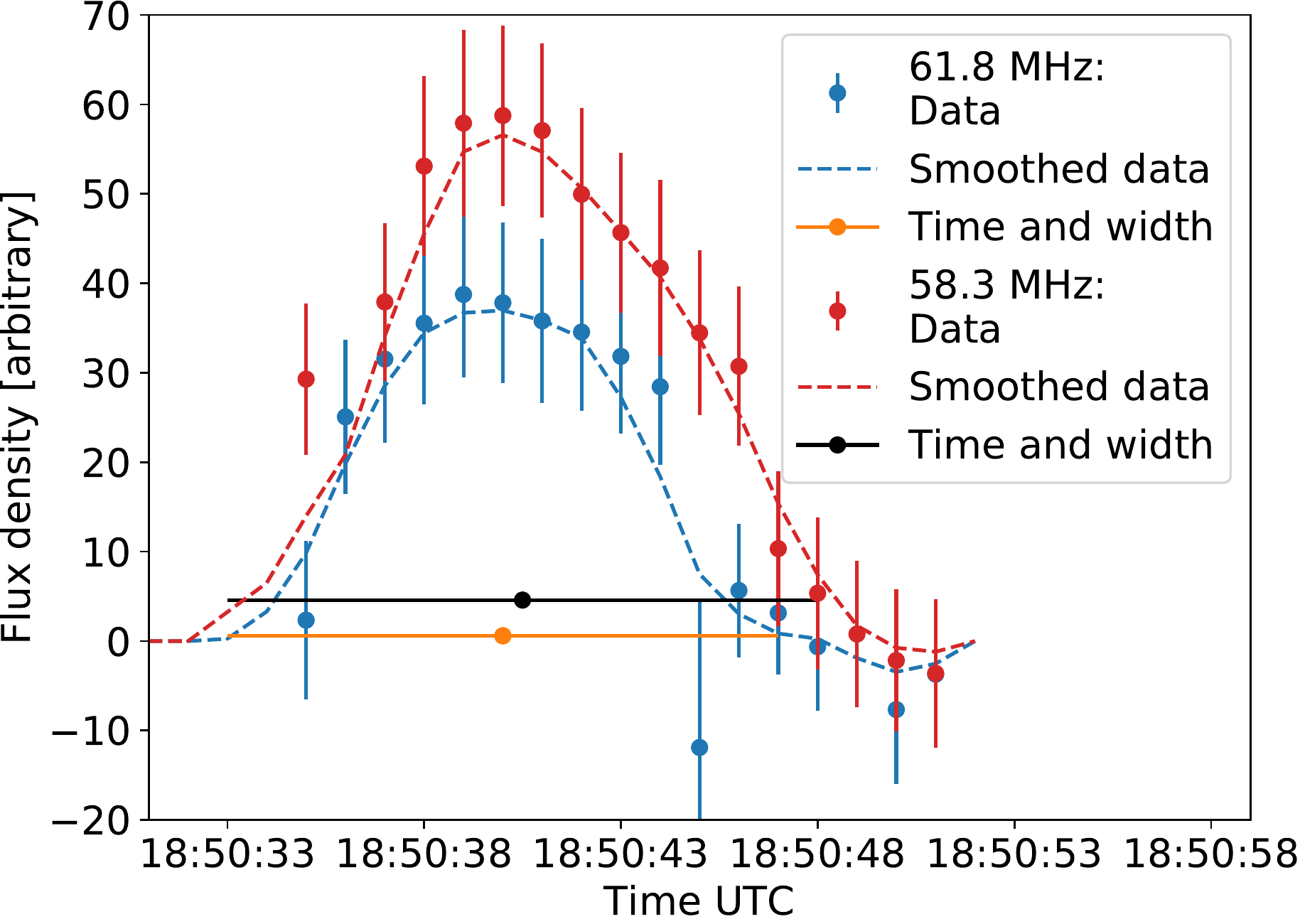}
\caption{An example light curve from an isolated burst-like magnification, observed on 2016-09-05. The blue and red circles depict the flux density measurements at 58.3 and 61.8 MHz, respectively (in arbitrary units, since the transient finder runs on the data before putting them on the correct flux scale). Dashed lines show the light curve smoothed over three seconds, used during peak analysis. The black and orange dots and horizontal lines show the peaks' mid-point, used for inter-frequency association, and the width at 10\% of peak flux level. The underlying source is 4C 26.55, a galaxy with a quiescent flux density of 15.9\,Jy at 74\.MHz \citep{2014MNRAS.440..327L}.}
\label{fig:TR1_discovery}
\end{figure}

Using this feature detection algorithm, we searched the candidate databases, output by TraP, for light curves consisting of only a single burst. 
Figure  \ref{fig:TR1_discovery} depicts an example transient candidate consisting of one burst, and the various parameters of interest described above.

\subsection{Strong scintillation zones}

We sometimes find large concentrations of magnification events in sky region and time, which we call
`strong scintillation zones'. We quantify these by counting the number of sources whose flux density is typically below the confusion noise limit, rendering them undetectable during calmer conditions, but are temporarily magnified above our threshold. 
This is done by extracting all of the sources observed in each 5 minute interval, and calculating the surface density of sources which are not present in the AARTFAAC catalogue. 
We model the source distribution in the sky with a Gaussian kernel density estimation (KDE), which yields a 2D probability density function (PDF) of detected sources. 
By comparing relatively calm observations to those where a zone is plainly visible by eye, in time-lapse videos, we heuristically set the PDF threshold which is characteristic of strong scintillation.

\section{Results}
\label{sec:results}

Here we report on three different types of transient or variable events observed in the seconds-timescale, low-frequency radio sky. 
The first are  isolated, spontaneous flashes, which coincide in position with 10--20\,Jy radio sources,  that are otherwise undetectable by AARTFAAC. Although they are coincident in  direction,  the flashes  do not have a sufficient time-frequency delay to originate from an extragalactic distance. 
The second, is strong variability characterised in some sources by many magnification spikes in a relatively short time, and in others by singular flashes similar to the first, both clearly induced by a localized structure  which persists over many hours. 
And third, we observed a few potential cosmic transients which, although similar to the other transients in peak shape, duration, and magnitude, have a
time-frequency delay that is consistent with an extragalactic origin, these sources are analysed further in the AARTFAAC low-frequency transient survey paper \cite{2020arXiv200313289K}.

\subsection{A detailed look at light curves}
\subsubsection{Singular magnification events}
\label{sec:singlemag}

\begin{figure*}
\includegraphics[width=\textwidth]{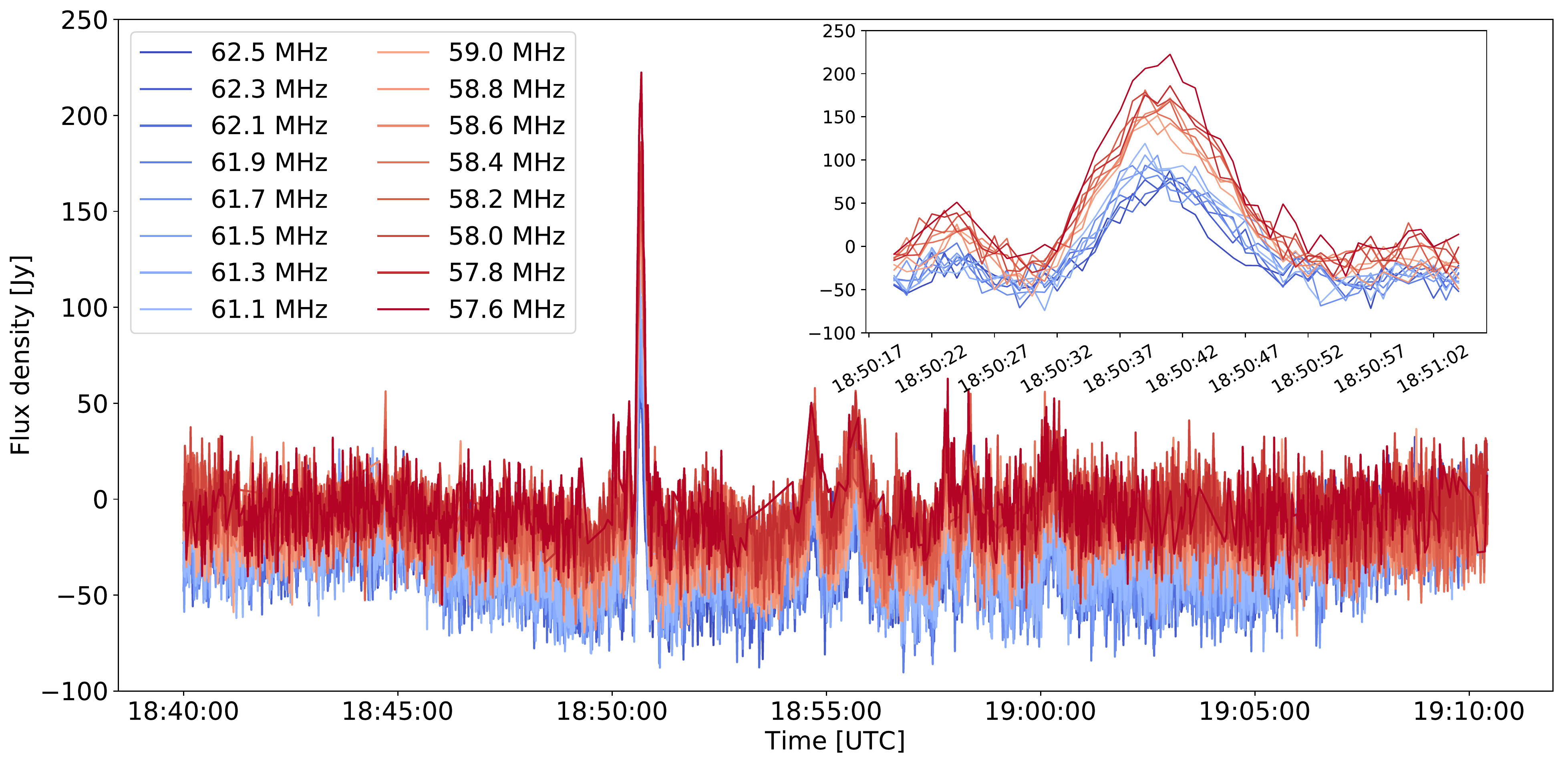}
\caption{Singular non dispersed burst-like profile from the 16 frequency channels sample with a 1 second time resolution, observed on 2016-09-05.}
\label{fig:TR1_lightcurve}
\end{figure*}

Using the method described in Section \ref{sec:analysis}, a population of 76 transient candidates, each of which appears as a single isolated burst, were discovered. 
One example observed on 2016-09-05, is shown in Fig. \ref{fig:TR1_discovery}. This light curve shows a magnification of the flux density of 4C\,26.55, which has a quiescent flux density of 15.9\,Jy at 74\,MHz \citep{2014MNRAS.440..327L}. This burst 
was  analysed more deeply by re-imaging all 16 individual sub-bands in a 30 minute interval around the candidate. 
The resulting 16 sub-band light curve is shown in Fig. \ref{fig:TR1_lightcurve}. Here, we can see the relative stability of the background in the 10 minutes prior, and 20 minutes after that burst, albeit with a few sub-detection-threshhold correlated noise events around 18:55. This indicates that this region of sky, at this time, is not subject to the type of strong scintillation described in Section \ref{sec:stats}. 

An isolated, bright outburst, with a time scale on the order of seconds is exactly the kind of transient signal we are searching for with AARTFAAC. This could potentially indicate the detection of a low-frequency fast radio burst (FRB), the prompt emission of a neutron star merger, or another as of yet unobserved phenomenon dumping a tremendous amount of energy in to coherent emission in a short time, an exciting prospect.
However, given the lack of any noticeable time delay between 62.5 and 57.6 MHz, it is apparent that the signal can not be of extragalactic origin. In Fig.~\ref{fig:TR1_lightcurve}  we see that there is no appreciable delay between the sub-bands. 

\begin{figure}
\includegraphics[width=\columnwidth]{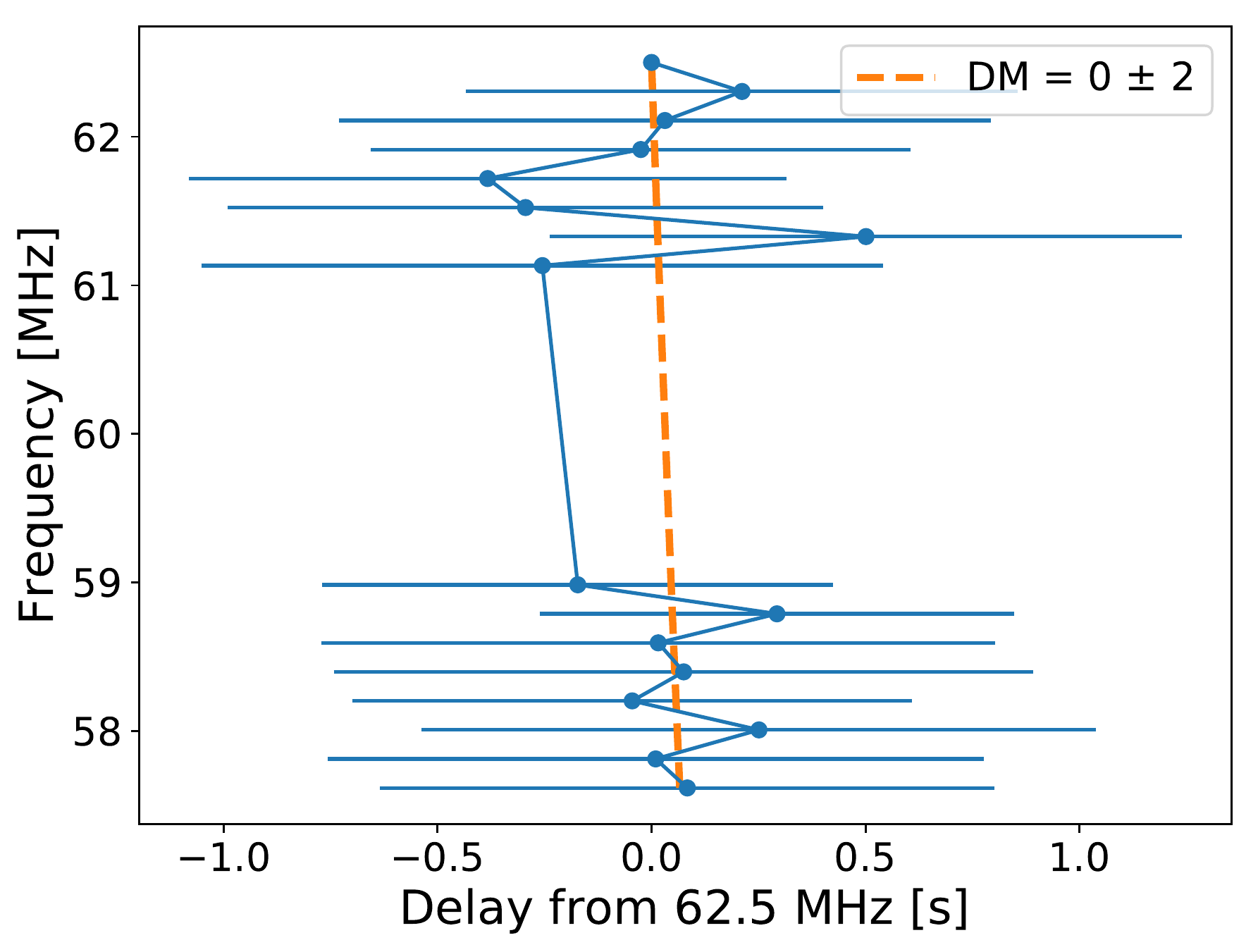}
\caption{The time lag between the 62.5 MHz light curve, the highest sub-band frequency, and each of the 16 sub-bands.  The delay uncertainties were calculated by randomly generating 10,000 sets of light curves using the measured flux densities and their uncertainties. The orange line shows the best fit dispersion delay, and is clearly consistent with 0 DM. Additionally, with the data quality during this event, we are sensitive to DMs of even a few.   }
\label{fig:DM_measure}
\end{figure}

In order to estimate an upper limit on the distance we attempted to measure any dispersion delay in the signal.
The time delays are measured by the lag in the cross-correlation functions between the light curve at 62.5 MHz and each successive frequency.
Additionally, in order to estimate the uncertainty in the inferred DM, we use a simple Monte-Carlo method: 
We randomly generate 10,000 sets of 16 light curves, according to the measured flux densities and uncertainty, at each frequency and time.
We then fit the delays for each set. The mean and standard deviation for each frequency are shown in Fig. \ref{fig:DM_measure} and allow us to conclude that
for this transient candidate, the DM is $0 \pm 2$, indeed consistent with zero.

\begin{figure}
\includegraphics[width=\columnwidth]{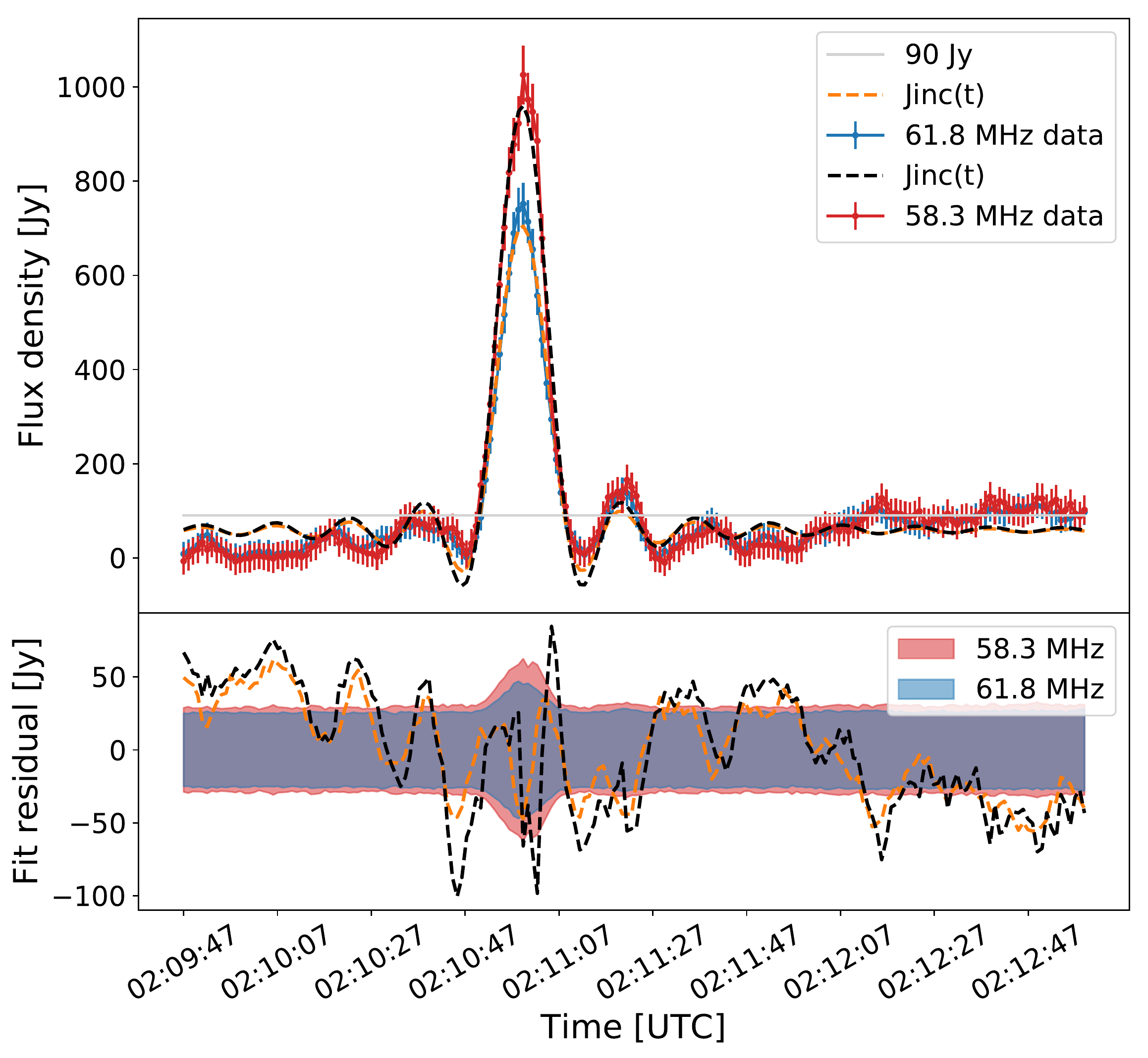}
\caption{Isolated magnification of the signal from 3C\,234, captured during 2019-01-13.  We fit a Jinc function (top), and plot the residuals (model $-$ data, bottom), where the shaded regions are defined by the flux density uncertainty. The horizontal grey line in the top panel indicates the 90\,Jy persistent flux density as measured in the AARTFAAC catalogue.  }
\label{fig:jinc_fit}
\end{figure}

In Fig.~\ref{fig:jinc_fit} we illustrate another example of an isolated,  non-dispersed burst-like event. Here, the magnified source is 3C 234. With a persistent flux density of $90.06 \pm 0.15$ Jy during regular conditions \citep{10.1093/mnras/sty2810}, the light curve surrounding the central magnification remains above the detection threshold. This reveals how the flux density is de-magnified, prior to, and following the apparent burst, before returning to the steady state. Together with some secondary peaks, it somewhat resembles
 an Airy function, except that the secondary peaks are a bit high relative to the central peak. We fit a Jinc function, which fits both the central peak and the positions and amplitudes of the satellite peaks fairly well:
\begin{equation}
    \mathrm{Jinc}(t) = a\frac{2 J_{\mathrm{1}}(t)}{t}+d,
\end{equation}
where $J_{\mathrm{1}}(t)$ is the Bessel function of the first kind, and  $t = \frac{x}{b} - c$, with $b$ and $c$ used to scale and translate the time $x$ of the light curve during fitting.
This pattern could be caused by a cone-shaped notch in a plasma sheet, but we shall return to that in the discussion (Sect.~\ref{sec:singlenodm}).

\subsubsection{Strong variability due to scintillation}
\label{sec:scintillation}

\begin{figure*}
\includegraphics[width=\textwidth]{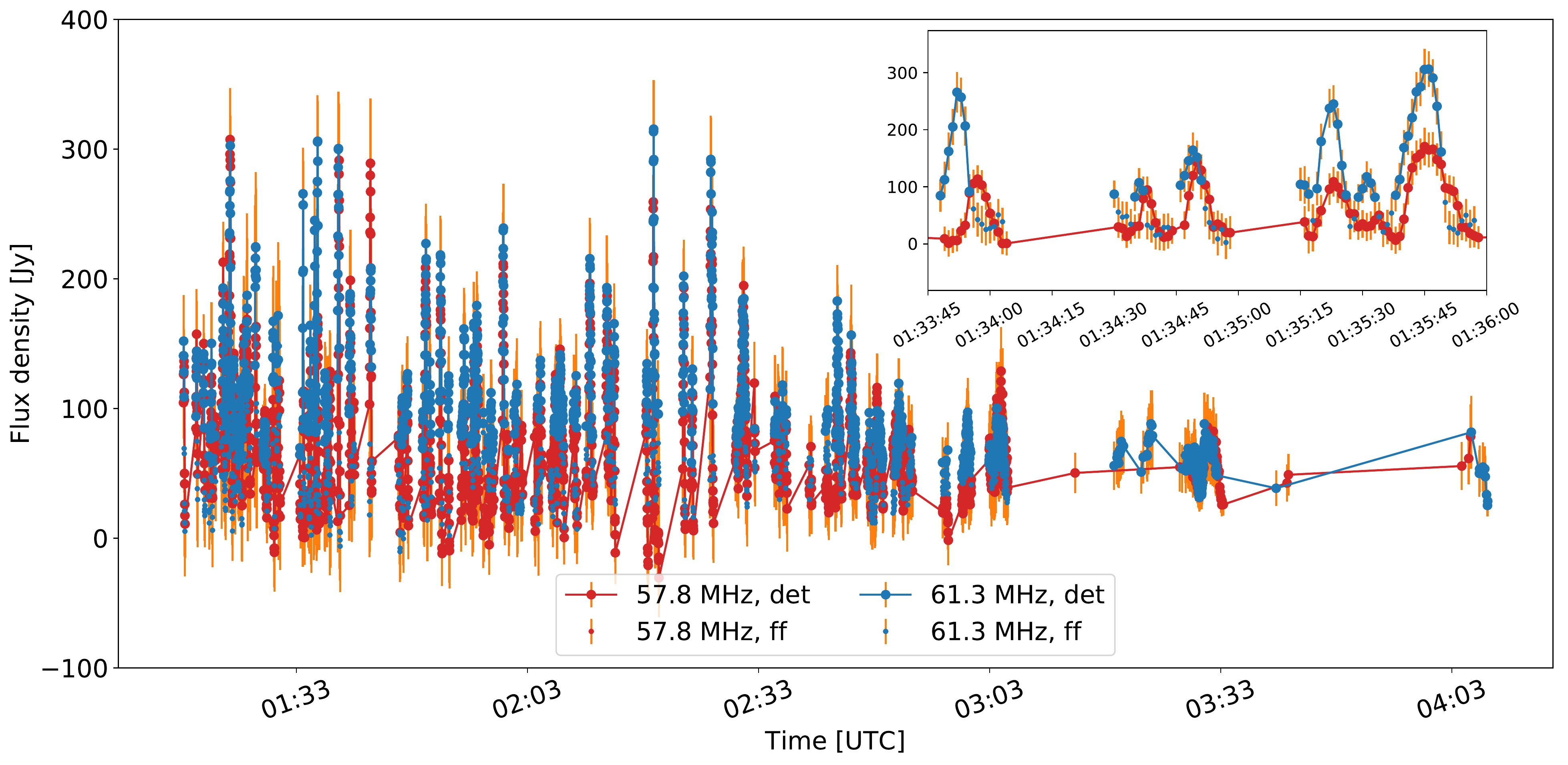}
\caption{Example light curve illustrating strong scintillation of 4C\,14.27 during an observation on 2019-01-01, with the inset expanding a typical region of the light curve, illustrating the scintel shape and height variability. The quiescent flux is 15.07\,Jy at 74\,MHz, typically below our detection threshold.  Large connected dots labelled ``det'' denote measurements where the signal is blindly detected, i.e., above $5\sigma$. smaller dots labelled ``ff'' are forced-fit fluxes, which are done automatically for a few time steps after a detection at the same location.} 
\label{fig:scintel_lightcurve}
\end{figure*}

Sometimes events similar to those described in the previous section appear in large
numbers in a region of the sky, affecting a given source multiple times during some period
and affecting many neighbouring sources as well. 
Figure \ref{fig:scintel_lightcurve} is an example light curve, recorded on 2019-01-01, when such strong scintillation was observed across nearly half of the visible sky.  
The source is 4C 14.27, a Seyfert 1 galaxy, with a flux density of  $15.07 \pm 0.16$ Jy in the VLSSr catalogue, and very compact, <10" according to the FIRST survey, which has a synthesized PSF FWHM of 5"  \citep{2015ApJ...801...26H}. 
We will call the individual events `scintels', and during such periods of strong scintillation, many tens of scintels, each similar in character to the isolated case in Fig. \ref{fig:TR1_lightcurve}, can be produced. 
This is clearly a persistent background source temporarily exhibiting strong variability through scintillation. 
In Section \ref{sec:stats}, we will measure the properties of these scintels and compare their distributions with the distributions of properties of the isolated events.

\subsection{Statistical description of strong scintillation zones}
\label{sec:stats}

\begin{figure*}
\includegraphics[width=\textwidth]{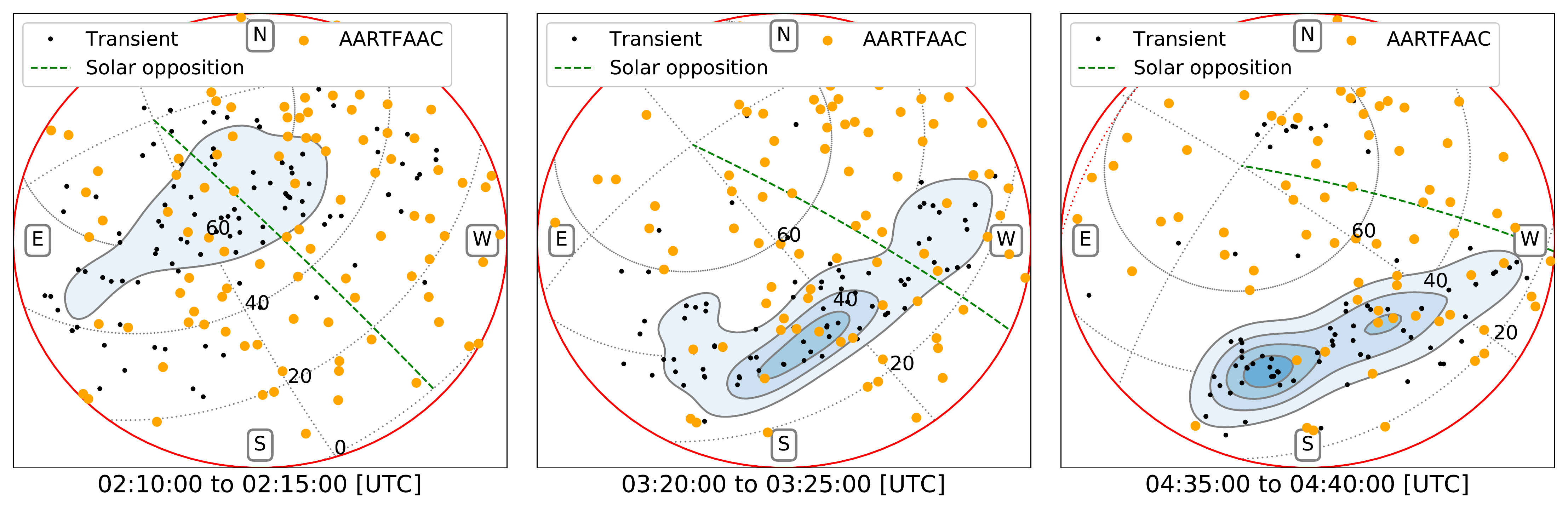}
\caption{Evolution of the scintillation zone, during three hours on 2017-02-25. Each frame illustrates the source detection region, down to 55 degrees from zenith (red line). Ecliptic coordinates (grey dotted lines), with the latitudinal parallels labelled at $180^{\circ}$ longitude. The green dashed line defines geomagnetic midnight, the longitude of solar opposition.  The AARTFAAC catalogue sources (orange circles), and other sources detected (black dots) during each time interval are shown. The contours outline the KDE of the scintels, with levels from 0.15--0.35, in steps of 0.05,  $\rm scintels~deg^{-2}~5~min^{-1}$  and reveal the extent of the zone of strong scintillation.  }
\label{fig:strip_evol}
\end{figure*}

In attempting to disentangle the different scintillation regimes discussed earlier, we have observed zones of strong scintillation  in opposition to the Sun.
Their locations are not generally fixed either in celestial or terrestrial coordinates, so it is not immediately clear from their motion what to associated them with. 
An example of these zones is shown in Fig. \ref{fig:strip_evol}\footnote{For an animated version visit \url{https://aartfaac.org/images/scint_20170225.gif}}.
The zones vary in size from 20 to 40 degrees ecliptic latitudinal extent and from 20 to 90 degrees ecliptic longitudinal extent. 
They can appear stationary in ecliptic coordinates and in size, or evolve along the latitude lines, decreasing from 60 to 40 degrees latitude, or remain stable between 0 and 20. 
Of the 28 nights where we observed between 22 and 05 UTC, strong scintillation regions are clearly defined in seven. 
The total lifetime of the zone can be hours, as seen in the observation on 2018-03-09, where the scintillation zone begins around 22:00 UTC, peaks in intensity at 00:00 UTC, then diminishes until disappearing at 04:00 UTC, lasting a total of 6 hours.  
However, on 2017-02-25, Fig. \ref{fig:strip_evol}, the zone is present at the start of the observation at 1:30 UTC, increases in intensity until 4:20 and then begins to decrease until the end of the observation at 5:00.  
Then in the next  night at 1:30, when we start observing again, the zone is present at 40 degrees latitude, but restricted in longitudinal extent compared with the previous night, indicating that this may be a single case lasting more than 24\,h.

\subsubsection{Magnification factor and peak flux of the underlying source}
\label{sec:stats_mag}

\begin{figure}
\includegraphics[width=\columnwidth]{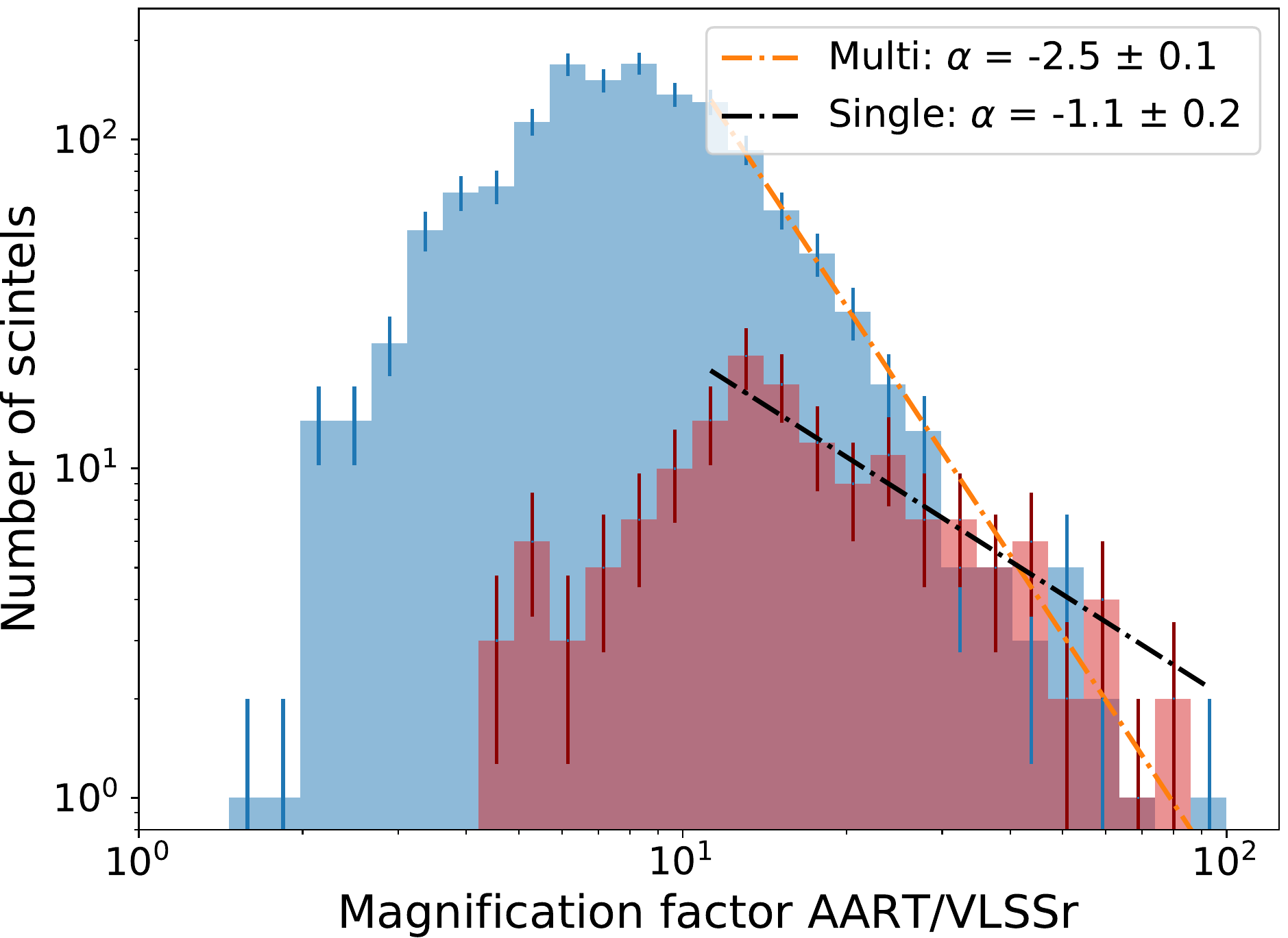}
\caption{Magnification factor distributions for the 725 multiply-magnified scintels (blue), and the 76 singly magnified scintels (red).}
\label{fig:scintel_mag_dist}
\end{figure}

In an effort to understand whether the singular magnification events are different in cause and character to the scintels observed during periods of strong scintillation, we compare the magnification distributions from multiply and singly magnified sources.
In Fig. \ref{fig:scintel_mag_dist} we show the distribution of magnification factors of 725 scintels described in Section~\ref{sec:scintillation}. 
These were measured from 16 sources, during one period of strong scintillation.  
The typical flux density of each is below our detection threshold, thus the individual scintels are well isolated from each other. 
Additionally,  76 singular scintels, instances where the flux density has been magnified only once during the observation (Section \ref{sec:singlemag}),  for that specific  background source, are also plotted. 
In both populations, the distribution of magnification factors is bounded on the lower side by our sensitivity limit. 
Therefore, the bottom of the distribution follows a power law, with index $-1$.  
The distribution is bounded on the top by the decreasing likelihood of observing a scintel with greater and greater magnification, and on the right by the decreasing number of sources at higher unmagnified brightness.
The distribution of magnification factors, illustrated in Fig. \ref{fig:scintel_mag_dist}, drops according to a power-law distribution with an index of $-2.5 \pm 0.1$ for the multiply magnified scintels. 
For the singly magnified scintels the distribution is flatter, with a power-law index is  $-1.3 \pm 0.2$.

\subsubsection{Duration distribution}
\label{sec:stats_duration}

\begin{figure}
\includegraphics[width=\columnwidth]{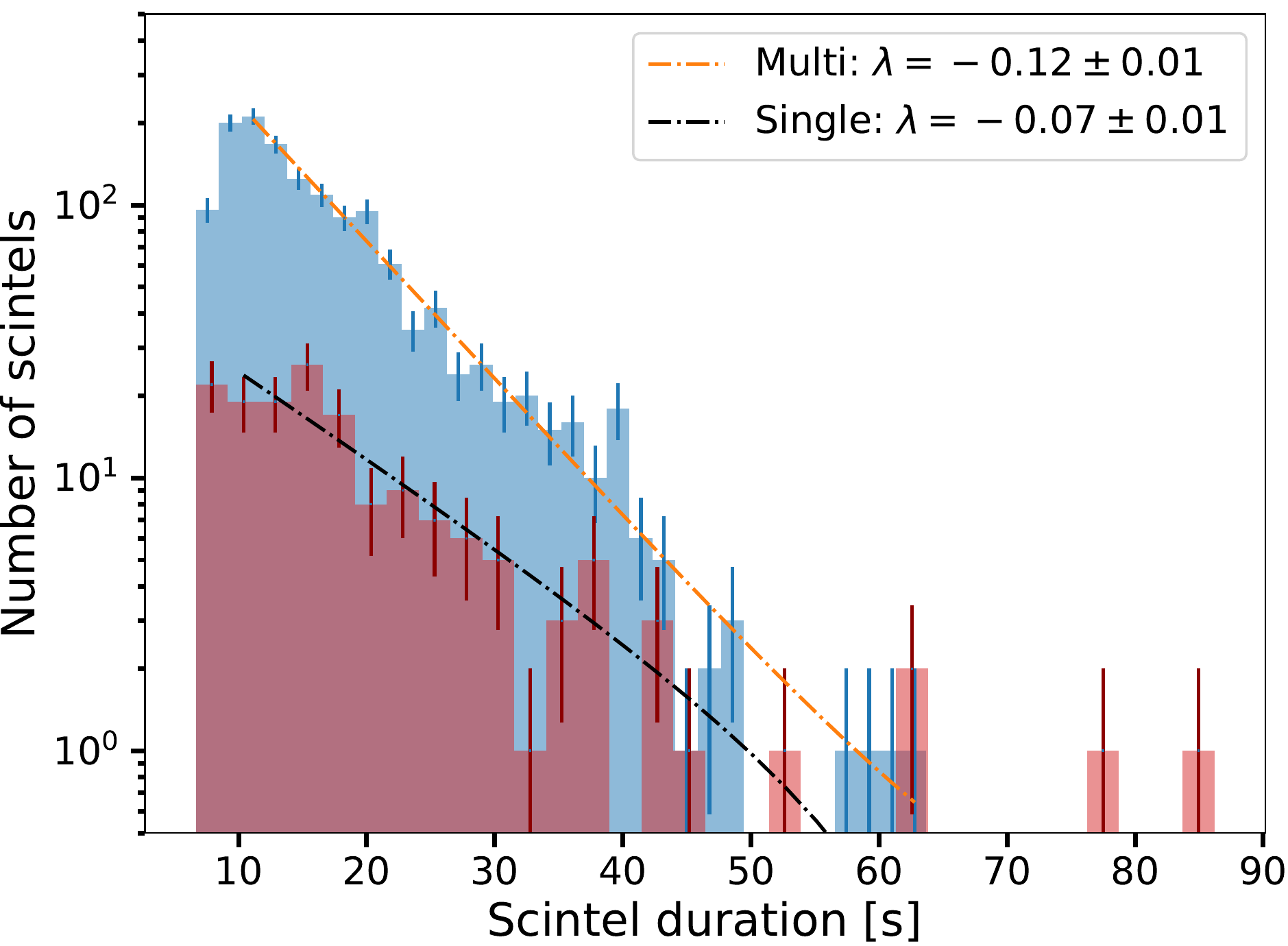}
\caption{Distribution of scintel duration for multiply magnified sources (blue) and single events (red).  closely follows an exponential distribution. }
\label{fig:scintel_width_dist}
\end{figure}

\begin{figure}
\includegraphics[width=\columnwidth]{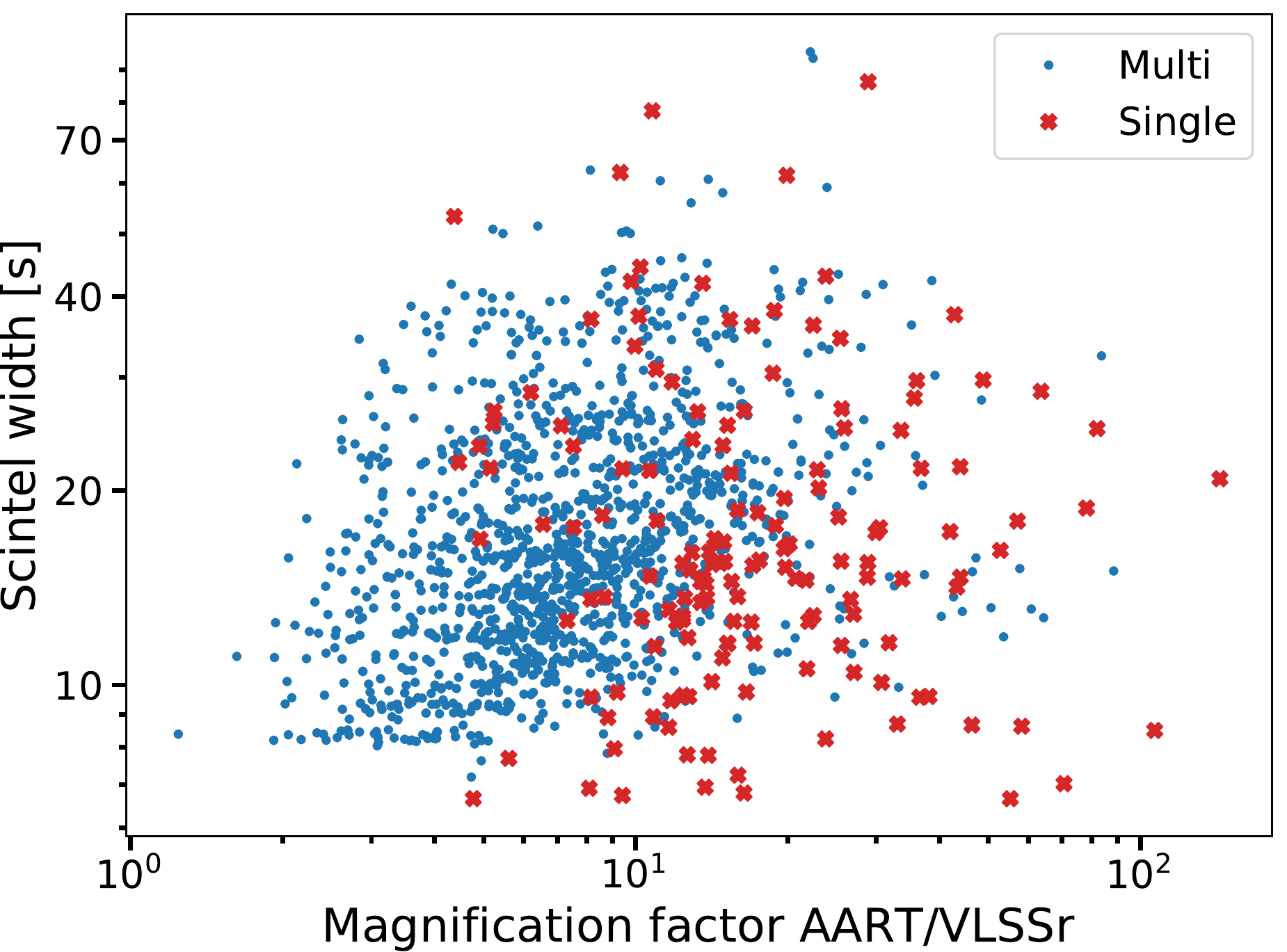}
\caption{Duration vs. magnification. The quantisation of the width is due to our one second sampling time.}
\label{fig:scintel_mag-width}
\end{figure}

Next, in an effort to infer physical properties of the turbulent medium producing the scintillation, we investigated the distribution in the temporal widths of the scintels, as well as the relationship between the scintels' width and the peak magnification factor. 
The width distribution is approximately exponential with a characteristic time of
 of $\tau=1/\lambda = 8.5 \pm 1.5$\,s, from the peak of 10\,s (Fig. \ref{fig:scintel_width_dist}). For comparison, the isolated magnification events have a somewhat longer characteristic duration, with a characteristic time of a fitted exponential distribution of $\tau= 15\pm 2$\,s.
Additionally, the scintel duration and magnification factor are almost uncorrelated
(correlation coefficient $0.2$, Fig.\ref{fig:scintel_mag-width}).

\subsubsection{Scintel delay distribution }
\label{sec:stats_delay}

\begin{figure}
\includegraphics[width=\columnwidth]{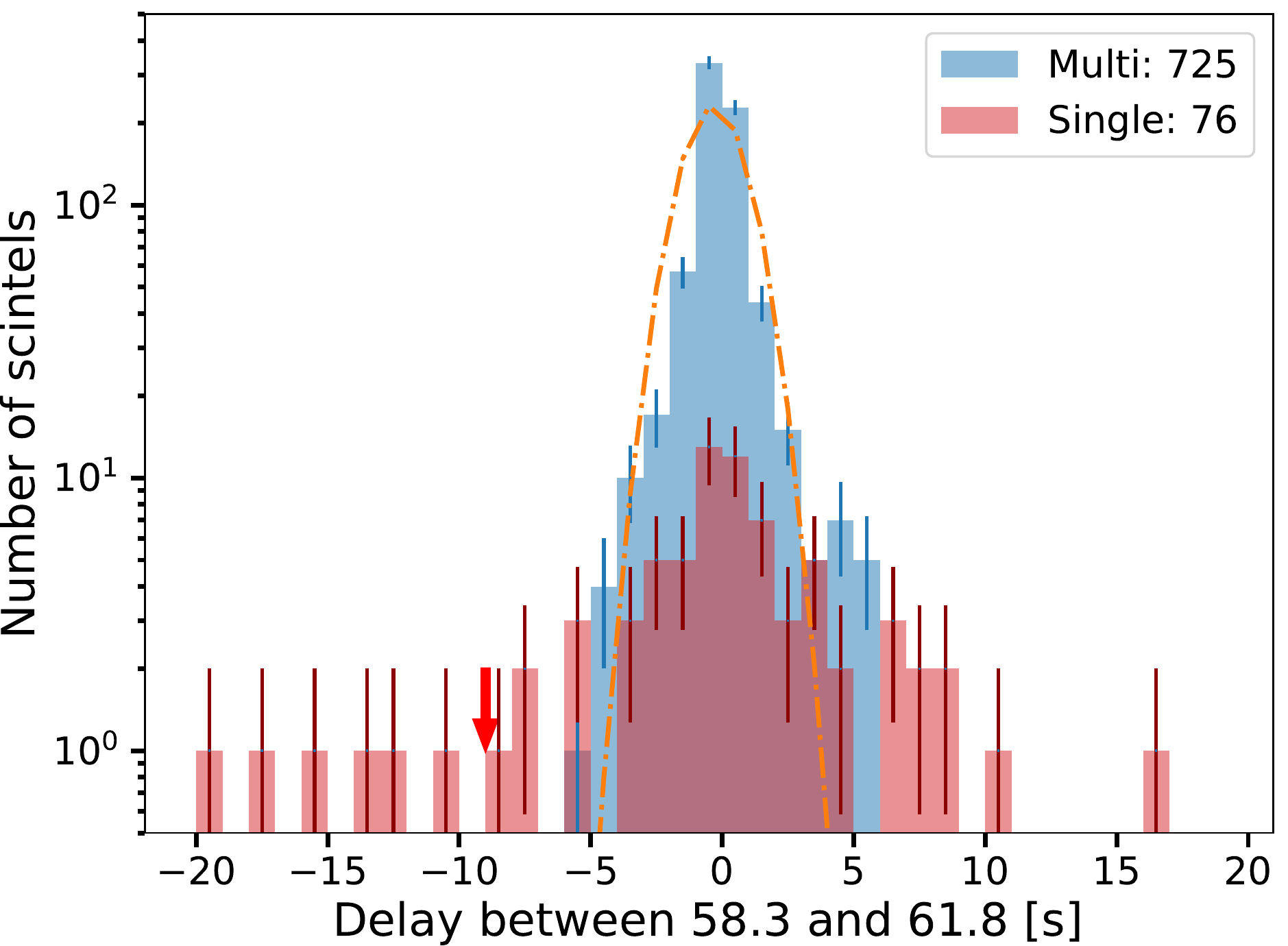}
\caption{Distribution of peak delay between 61.8 and 58.3\,MHz for 725 isolated scintels from 16 different sources on 2019-01-01 (blue), and 76. The red arrow denotes the delay of the transient candidate presented in Kuiack et al. 2020b. 
The delay is consistent with an extragalactic distance. The orange dash-dot line shows a Gaussian distribution fit to the population.}
\label{fig:delay_stats}
\end{figure}

Lastly, in order to deduce to what extent strong scintillation confounds the search for extragalactic transients, we investigated the distribution of delay times between peaks measured at 61.8 and 58.5\,MHz for each scintel or isolated magnification event. The distributions are shown in Fig. \ref{fig:delay_stats} for both types separately. The 
distribution of scintel delay times is fairly narrow and symmetric around zero, and its
width is consistent with noise errors. Hence we conclude that
they have no measurable delays and thus are consistent with DM$=0$.  For the isolated
events, the majority fall into a similarly narrow and symmetric distribution, but here
we find a significant number of outliers, and more so on the negative side where true dispersion delays should lie. The potential extragalactic event described in detail in \cite{2020arXiv200313289K} is indicated with the red arrow.

\section{Discussion}
\label{sec:discussion}

As is clear, most of the phenomena we described above are not consistent with a distant astrophysical origin and must be caused by plasma lensing or other propagation effects relatively near to Earth. We now discuss what constraints we can place on the medium that causes them. 

\subsection{Detailed light curves of single magnification events}
\label{sec:singlenodm}

Among the single magnification events, there are a few with very high signal to noise, and we consider these first in order to see what we can learn from the details of their light curves. The first light curve is that of 4C 26.55, shown in Fig. \ref{fig:TR1_lightcurve}.
The total DM through our Galaxy along this line of sight is $65 \mathrm{~pc~cm^{-3}}$ or  $71 \mathrm{~pc~cm^{-3}}$ according to the  YMW16 \citep{2017ApJ...835...29Y} and NE2001 \citep{2002astro.ph..7156C} models, respectively. 
Using only this range, which would ignore any contribution from the intergalactic medium or the local environment would result in a dispersion delay of 12.3--13.3\,s between 57.6 and 62.5\,MHz. However, we measure no significant delay, and our earlier analysis (Fig. \ref{fig:DM_measure}) showed a formal measured DM of $0\pm2$. Hence, the brightness
increase cannot originate in the source, but must be caused by plasma within a few tens of parsecs.

Our highest signal-to-noise magnification event peaks at 1000\,Jy  (Fig. \ref{fig:jinc_fit}). Again, the lack of measurable dispersion delay indicates the lensing plasma must be fairly nearby. A Bessel beam diffraction pattern, which matches this light curve, could be caused by a conic depression in a higher-density, plane parallel plasma sheet. (Since the index of refraction due to cold free electrons is close to,
but below unity, a converging lens must be under-dense and convex relative to its surroundings, or overdense and concave.) The somewhat depressed flux before and after the peak relative to the normal state may also suggest we see the `shadow' of this lensing plasma there.

One type of magnification event known to occur for compact sources
is the so-called Extreme Scattering Event (ESE) \citep[e.g.,]{1998ApJ...496..253C}, caused by isolated bubbles in the interstellar electron density. We do not think these are exactly the cause here,
because they would typically be further away and also could not cause the 
scintillation storms, in which the individual events look similar. They
also have time scales of months to years typically and amplitudes of only
a factor few. Furthermore, \cite{2018MNRAS.481.2685D} show that the lensed-to-unlensed brightness ratio tends strongly to one below 700 MHz, resulting from the increase in background source size. 

We also note that these singular magnifications are similar in shape, timescale, and magnitude (in the case of the burst in Fig. \ref{fig:jinc_fit}) to a recently reported low-frequency cosmic transient observed by LWA1 and LWA-SV \citep[][esp.\ Figs.~3 and 4]{2019ApJ...874..151V}. While this was not their preferred explanation,
they noted that it could be a magnification  of the
compact radio source 4C$+$1.06 by about a factor 30, a value on the high side, but within the range we find here; its duration of 15--20\,s is also within our range. The observing frequency of 34\,MHz is even lower than ours. If we may therefore regard this event as part of the same class the LWA gives important extra information: while they report only one event, and without dispersion information, the fact that they detected it with two stations 75\,km apart is important. First, the absence of a noticeable parallax between the two stations indicates a minimum distance of 1400\,km to the lens. LWA also report other events detected that they do attribute to scintillation (17 over a period of about 340 days). These are much less frequent than the scintillation storms we find, but that is likely due to the much higher flux threshold of LWA and the fact that they limited their analysis to cases where there was a simultaneous spike in the two stations. Due to this selection, we should probably be careful to attach too much significance to the simultaneity of these events in the two stations (to within about 5\,s). Such simultaneity would imply a size of the magnification spot on the ground of more than 75\,km, and a transverse speed of any lensing plasma of at least a few tens of km/s. However, if we look at their Fig.~4, we also see many more scintillation spikes in their example station light curves in the half hour before and after the simultaneous spike that are not coincident between the two stations, so smaller and/or slower magnification spots are perhaps more common. 
Before discussing a toy model for these magnifications, we will first look into what additional constraints follow from the strong scintillation zones. 

\subsection{Constraints from strong scintillation zones}

To the extent that we may regard the scintillation zones or periods as the occurrence of many events like the isolated ones in a short period of time, they provide at least one extra constraint: we can see them move in the sky and thus get their angular velocity. We use the 
example in Fig. \ref{fig:strip_evol}, where the apparent motion is from $60^\circ$ to $40^\circ$ ecliptic latitude in 70 minutes, i.e., an angular
velocity of $8\times10^{-5}$\,rad/s. 
First of all, this angular velocity allows us to very much reduce the 
maximum distance of the lensing plasma from the dispersion delay limit of
a few tens of pc: since the physical transverse speed must be
less than the speed of light, the distance is at most 20\,AU. Other than
that, the scintillation zone is clearly not
fixed either relative to the horizon or in ecliptic coordinates, so it cannot be obviously associated with either phenomena fixed to the Earth (such as the flux tubes discovered by \cite{2015GeoRL..42.3707L,2016JGRA..121.1569L} with the MWA), or with something fixed in the solar system. We do note a preference for the zones to appear
in the anti-Sun half of the sky, though, perhaps suggesting a link with downstream effects of the solar wind impacting on the Earth's magnetosphere.
To analyse this constraint in more detail, we must first note that the scintillation zones are tens of degrees across in the sky, but as we will show below, the individual lensing plasma bubbles are much smaller.  However, scintillation studies done with LOFAR by \cite{2020JSWSC..10...10F} indicate that a fair approximation is that a scattering screen with turbulent cells moves past our line of sight rather faster than that it changes in its own rest frame, as in the case of interstellar scintillation. Therefore we will assume that the angular velocity we measure for the overall region also applies to the individual turbulent cells.

In a recently published study of a strong scintillation episode in Cas\,A
with LOFAR \cite{2020JSWSC..10...10F} perform a detailed analysis of dynamic spectra and of the time sequence in which the scintillation hits different LOFAR stations. They find that this episode is caused primarily by two layers of the ionosphere,
the D region at around 80\,km, where the plasma speed was about 110\,m/s and the
F region at 300--400\,km altitude, where that speed was 20--40\,m/s at the
time. Furthermore, during a 31 hour survey with OVRO-LWA  between 27 and 84\,MHz \cite{2019ApJ...886..123A} observed source flux increases up to 100s of Jy with timescales on the order of 13 seconds (their highest time resolution), this is explained by a scattering screen at a distance of 300\,Km, again in the F region of the ionosphere. The typical quiescent electron densities in these regions are 0.5--1$\times10^6$\,cm$^{-3}$. For the angular velocity we measure, the
implied speeds would be 6\,m/s in the D region and 25--35\,m/s in the F region;
so in case we see the F region, we find about the same speed, but were it the
D region then the disturbance we see is considerably slower than the one found
by \cite{2020arXiv200304013F}. For even higher altitudes, as suggested by the lower limit for the LWA event, the required speeds go up: to 200\,m/s at an altitude of 2000\,km or even 1\,km/s at 1.5 Earth radii in the geomagnetic tail.

The range of  velocities, $100-1000$\,m/s, and altitudes ranging from $300-2000$ km matches well with travelling ionospheric disturbances  \citep[(TIDs, e.g.,]{1968JGR....73.6319T, 1998JGR...103.2131H}, as does the trajectory southward from high latitude \citep{1996AnGeo..14..917H}. The northwest to southeast orientation, and southwesterly drift is also characteristic of TIDs observed at mid-latitudes \citep{2000JGR...10518407G}. 
However, in the following hour the region remains fixed on the sky, along the $40^{\circ}$ parallel. 
\cite{1968JGR....73.6319T} also report a decrease in velocity from 700 m/s to 125 m/s over 11 hours. Lastly, direct air-glow images of  TIDs tend to show multiple waves spanning the entire sky, passing with motion uncorrelated to the background sidereal drift. 

Therefore we conclude that the information we get from the motion of the scintillation regions provides the severest distance restriction and confines the location of the scattering or lensing screen to within 20\,AU. The motion is consistent with a variety of plausible ionospheric or
geomagnetospheric origins and does not strongly favour one among these.

\subsection{A toy plasma bubble model}
\label{sec:toy}

We now try to constrain the properties of the plasma that causes both the isolated events and scintillation storms. To this end,
let us first assume a spherical bubble of radius $R$ with a difference in index of refraction $\delta n$ relative to its surroundings, that is hit by a plane wave from our source, which then converges on AARTFAAC. In order to achieve a large magnification factor, the distance of the bubble to our telescope must be similar to the focal length $f$ of the bubble. For $n\simeq1$, this is well approximated by $f=R/|2\delta n|$, irrespective of whether the lens is thick or thin in the lens makers' sense. (If we think of it as a spherical bubble, it should be underdense, but we can also imagine an overdense concave body with both size and surface curvature radius about equal to $R$). Using the standard expressions for the index of refraction and plasma frequency, we derive the following numerical expressions:
\begin{eqnarray}
    |\delta n| &=& 2.24\times 10^{-2} \delta n_{\rm e,6} \nu_{60} \\
    f          &=& 223 R_6 \delta n_{\rm e,6}^{-1} \nu_{60}^2 \,\,{\rm km}, \label{eq:f}
\end{eqnarray}
where $\delta n_{\rm e,6}$ is the electron density in units of $10^6$\,cm$^{-3}$, $R_6$ the lens radius in units of 10\,km, and $\nu_{60}=\nu/(60\,{\rm MHz})$. Here we have put some prejudice in the numbers by scaling them to values relevant for the quiescent ionosphere. For the ideal ray optics case, the magnification becomes infinite in the focus, but of course there it will always meet the diffraction limit. More realistically, AARTFAAC will cut through the beam some distance before or after the focus as the bubble drifts through the line of sight and throws a spot of radius $r$ onto the ground. The magnification is then simply $(R/r)^2$. The formal diffraction limit on the spot size is then of course $r_{\rm min}=f\theta_D/2=0.305\lambda f/R=0.61\lambda/\delta n$. Using the same numerical normalisations again, we find
\begin{eqnarray}
   r_{\rm min}  &=& 0.136 \delta n_{\rm e,6}^{-1} \nu_{60} \,\,{\rm km} \label{eq:rmin}\\
   M_{\rm max}  &=& 5.4\times10^3 R_6^2 \delta n_{\rm e,6}^2 \nu_{60}^{-2} \label{eq:Mmax}\\
   \frac{\Delta f}{f} &=& 3.4\times10^{-2} R_6^{-2} \delta n_{\rm e,6}^{-2} \nu_{60},
\end{eqnarray}
where $M_{\rm max}$ gives the diffraction limit to the peak magnification, and the depth of field, $\Delta f$, is the range of focal length over which the spot size and magnification are close to the diffraction limit. This shows that these extreme numbers will be very rare even in the ideal case, and of course one should note that realistic plasma bubbles are quite unlikely to be so smooth as to act as ideal lenses, and so rather larger spots and more modest magnifications are likely to be the rule. But the observed magnification factors (Fig.~\ref{fig:scintel_mag_dist}) can at least be reassuringly far below that theoretical maximum. Of course the duration $T$ of an  event is related to its actual spot size as $T=2r/v_{\rm L}$, but unfortunately via the unknown transverse speed $v_{\rm L}$ of the lens.

Another quantity that is accessible to observational testing is the angular size of the lenses on the sky, since large lenses would affect nearby sources in similar ways. The angular diameter on the sky is of course simply $\theta_{\rm L}=2R/f$, again approximating that AARTFAAC's distance to the lens is similar to the focal length, which is needed for a significant magnification. Numerically this works out to
\begin{equation}
    \theta_{\rm L}= \frac{2R}{f} = 4\delta n=5.1^\circ \delta n_{\rm e,6} \nu_{60}^{-2}. \label{eq:theta}
\end{equation}
This is quite sizeable and less dependent on parameters than some of the other quantities. Unfortunately it is smaller than the typical distance between AARTFAAC catalogue sources on the sky, and so the test is non-trivial.

Now we can use these equations with the properties of the observed events and of AARTFAAC to derive some general constraints on possible lenses. First, we note that a detectable magnification will require a spot that covers the AARTFAAC superterp, i.e., $r_{\rm min}>300$\,m. Via eq.~\ref{eq:rmin}, this places a lower limit on the electron density pertubation: $\delta n_{\rm e,6}<0.45$. If we now use eq.~\ref{eq:f} and the requirement that $f\simeq h$, where $h$ is the altitude of the lens, we get that
\begin{equation}
    R_6 = \frac{h}{223\,{\rm km}} \delta n_{\rm e,6}.
\end{equation}
Here and below we set $\nu_{60}=1$, all observations reported here are near 60\,MHz. We can now set a constraint on the magnification by eliminating $R_6$ from it and using the electron density constraint:
\begin{equation}
    M_{\rm max} = 5.4\times10^3 \left(\frac{h}{223\,{\rm km}}\right)^2 \delta n_{\rm e,6}^4 < 230 \left(\frac{h}{223\,{\rm km}}\right)^2.
\end{equation}
Since we observe magnifications up to 120, this implies that some of them must originate from a height at least 160\,km, excluding the D region we mentioned previously. This exclusion is even stronger if we remind ourselves that we calculated the spot sizes and magnifications for ideal lenses, and realistically we therefore expect higher altitudes. Near the density peak of the F region, at perhaps 400\,km, one could theoretically get magnifications of 750 and perhaps that makes the observed values possible even with somewhat non-ideal lenses. In this layer the quiescent electron density is also highest, so that the maximum density perturbation we derive is only about half the total value. If one goes up to one Earth radius, the density has dropped by a factor 1000, more than undoing any favourable effect of a larger distance on the possible magnification. This is valid as long as we consider waves and turbulence that perturb the normal plasma values around the mean. If one considers events that change the electron density by orders of magnitude there would be more freedom.

We also note that for the spot sizes needed for large magnifications and the speeds we measure, the expected durations of the magnification events are a few seconds to a few minutes, which nicely encompasses the range we observed. From eq.~\ref{eq:theta} we also find that the angular size of the lenses must be less than 2$^\circ$. We find very few cases of sources closer than that on the sky that both scintillate, so with our current data we cannot test whether sources closer than that vary in sync because they are lensed by the same bubbles.

Hence indeed, taking all constraints together, we conclude that the layers around the ionospheric density peak at a height of about 300--400\,km are a plausible location for the plasma that causes the spiky magnification events that we have found. A more common description of the radio scintillation caused by  these layers considers the layer as a turbulent scattering screen that causes a Fresnel-zone near-field pattern of intensity variations on the ground. This approach is used in the study of scintillation of Cas\,A with LOFAR \citep{2020arXiv200304013F}. Quite possibly both descriptions apply simultaneously: Cas\,A is very bright, and a single LOFAR station can detect it, and its variations, continuously. Therefore monitoring it can reveal the continuous pattern cast on the ground by the turbulent layer. We here see only the tip of the iceberg, because all we detect is the occasional very large magnification due a fortuitously shaped turbulent cell and favourable geometry. We find that such events are much more common when the ionosphere is more active, but they can also occur in an otherwise quiescent ionosphere, and we find that the distributions of their properties are somewhat different then.

\section{Conclusions}
\label{sec:conclusion}

We have identified a new phenomenon in the low-frequency radio sky, namely strong and well-separated radio flashes, typically lasting 20\,s and mostly coincident with bright radio sources that become 3--30 times brighter during the flash, but occasionally up to 100 times brighter. They typically show no significant dispersion delays and thus cannot originate from intrinsic variations in the extragalactic sources they coincide with, so we interpret them as magnification events caused by near-Earth plasma. Using a naive toy model of a plasma lens, we show that they are consistent with originating in the ionosphere around the layer of peak electron density, 300--400\,km in altitude, so they may be the extreme end of ionospheric scintillation. They do occur both in times when the ionosphere is otherwise quiescent, albeit rarely, and they become quite frequent when the ionosphere is very active. At those times, which last from a few hours to more than a day, there are so many events that we can trace the development and motion of the active regions across the sky. In the future, with somewhat higher sensitivity and therefore a higher source density on the sky, we should be able to detect or exclude correlated variability of source that are less than 2$^\circ$ apart in the sky and thereby constrain the sizes and motion of the individual lensing plasma bubbles.

\section*{Data Availability}

The data underlying this article can be shared on reasonable request to the corresponding author.

\section*{Acknowledgements}

AARTFAAC development and construction was funded by the ERC under the Advanced Investigator grant no. 247295 awarded to Prof. Ralph Wijers, University of Amsterdam; This work was funded by the Netherlands Organisation for Scientific Research under grant no. 184.033.109. We thank The Netherlands Institute  for Radio Astronomy (ASTRON) for support provided in carrying out the commissioning observations. AARTFAAC is maintained and operated jointly by ASTRON and the University of Amsterdam.

We would also like to thank the LOFAR science support for their assistance in obtaining and processing the data used in this work. We use data obtained from LOFAR, the Low Frequency Array designed and constructed by ASTRON, which has facilities in several countries, that are owned by various parties (each with their own funding sources), and that are collectively operated by the International LOFAR Telescope (ILT) foundation under a joint scientific policy. We thank Harish Vedantham for useful discussions on the lensing mechanism and properties.

This research made use of Astropy,\footnote{http://www.astropy.org} a community-developed core Python package for Astronomy\citep{astropy:2013, astropy:2018}, as well as the following:  KERN \citep{molenaar2018kern}, Pandas \citep{mckinney-proc-scipy-2010}, NumPy \citep{2011CSE....13b..22V}, and SciPy \citep{citescipy}. Accordingly, we would like to thank the scientific software development community, without whom this work would not be possible.




\bibliographystyle{mnras}
\bibliography{ApparentTransients} 

\newcommand{\noop}[1]{}
\begin{thebibliography}{}
\makeatletter
\relax
\def\mn@urlcharsother{\let\do\@makeother \do\$\do\&\do\#\do\^\do\_\do\%\do\~}
\def\mn@doi{\begingroup\mn@urlcharsother \@ifnextchar [ {\mn@doi@}
  {\mn@doi@[]}}
\def\mn@doi@[#1]#2{\def\@tempa{#1}\ifx\@tempa\@empty \href
  {http://dx.doi.org/#2} {doi:#2}\else \href {http://dx.doi.org/#2} {#1}\fi
  \endgroup}
\def\mn@eprint#1#2{\mn@eprint@#1:#2::\@nil}
\def\mn@eprint@arXiv#1{\href {http://arxiv.org/abs/#1} {{\tt arXiv:#1}}}
\def\mn@eprint@dblp#1{\href {http://dblp.uni-trier.de/rec/bibtex/#1.xml}
  {dblp:#1}}
\def\mn@eprint@#1:#2:#3:#4\@nil{\def\@tempa {#1}\def\@tempb {#2}\def\@tempc
  {#3}\ifx \@tempc \@empty \let \@tempc \@tempb \let \@tempb \@tempa \fi \ifx
  \@tempb \@empty \def\@tempb {arXiv}\fi \@ifundefined
  {mn@eprint@\@tempb}{\@tempb:\@tempc}{\expandafter \expandafter \csname
  mn@eprint@\@tempb\endcsname \expandafter{\@tempc}}}

\bibitem[\protect\citeauthoryear{{Anderson} et~al.,}{{Anderson}
  et~al.}{2019}]{2019ApJ...886..123A}
{Anderson} M.~M.,  et~al., 2019, \mn@doi [\apj] {10.3847/1538-4357/ab4f87},
  \href {https://ui.adsabs.harvard.edu/abs/2019ApJ...886..123A} {886, 123}

\bibitem[\protect\citeauthoryear{{Astropy Collaboration} et~al.,}{{Astropy
  Collaboration} et~al.}{2013}]{astropy:2013}
{Astropy Collaboration} et~al., 2013, \mn@doi [\aap]
  {10.1051/0004-6361/201322068}, \href
  {http://adsabs.harvard.edu/abs/2013A%26A...558A..33A} {558, A33}

\bibitem[\protect\citeauthoryear{{Bell} et~al.,}{{Bell}
  et~al.}{2014}]{2014MNRAS.438..352B}
{Bell} M.~E.,  et~al., 2014, \mn@doi [\mnras] {10.1093/mnras/stt2200}, \href
  {http://adsabs.harvard.edu/abs/2014MNRAS.438..352B} {438, 352}

\bibitem[\protect\citeauthoryear{{Bell} et~al.,}{{Bell}
  et~al.}{2019}]{2019MNRAS.482.2484B}
{Bell} M.~E.,  et~al., 2019, \mn@doi [\mnras] {10.1093/mnras/sty2801}, \href
  {https://ui.adsabs.harvard.edu/abs/2019MNRAS.482.2484B} {482, 2484}

\bibitem[\protect\citeauthoryear{{Blagoveshchenskii}}{{Blagoveshchenskii}}{2013}]{2013Ge&Ae..53..275B}
{Blagoveshchenskii} D.~V.,  2013, \mn@doi [Geomagnetism and Aeronomy]
  {10.1134/S0016793213030031}, \href
  {https://ui.adsabs.harvard.edu/abs/2013Ge&Ae..53..275B} {53, 275}

\bibitem[\protect\citeauthoryear{{Carbone} et~al.,}{{Carbone}
  et~al.}{2016}]{2016MNRAS.459.3161C}
{Carbone} D.,  et~al., 2016, \mn@doi [\mnras] {10.1093/mnras/stw539}, \href
  {https://ui.adsabs.harvard.edu/#abs/2016MNRAS.459.3161C} {459, 3161}

\bibitem[\protect\citeauthoryear{{Clegg}, {Fey}  \& {Lazio}}{{Clegg}
  et~al.}{1998}]{1998ApJ...496..253C}
{Clegg} A.~W.,  {Fey} A.~L.,   {Lazio} T. J.~W.,  1998, \mn@doi [\apj]
  {10.1086/305344}, \href
  {https://ui.adsabs.harvard.edu/abs/1998ApJ...496..253C} {496, 253}

\bibitem[\protect\citeauthoryear{{Cordes} \& {Lazio}}{{Cordes} \&
  {Lazio}}{2002}]{2002astro.ph..7156C}
{Cordes} J.~M.,  {Lazio} T.~J.~W.,  2002, ArXiv Astrophysics e-prints, \href
  {http://adsabs.harvard.edu/abs/2002astro.ph..7156C} {}

\bibitem[\protect\citeauthoryear{{Danilov} \& {Konstantinova}}{{Danilov} \&
  {Konstantinova}}{2019}]{2019Ge&Ae..59..554D}
{Danilov} A.~D.,  {Konstantinova} A.~V.,  2019, \mn@doi [Geomagnetism and
  Aeronomy] {10.1134/S0016793219050025}, \href
  {https://ui.adsabs.harvard.edu/abs/2019Ge&Ae..59..554D} {59, 554}

\bibitem[\protect\citeauthoryear{{Dong}, {Petropoulou}  \& {Giannios}}{{Dong}
  et~al.}{2018}]{2018MNRAS.481.2685D}
{Dong} L.,  {Petropoulou} M.,   {Giannios} D.,  2018, \mn@doi [\mnras]
  {10.1093/mnras/sty2427}, \href
  {https://ui.adsabs.harvard.edu/abs/2018MNRAS.481.2685D} {481, 2685}

\bibitem[\protect\citeauthoryear{{Fallows} et~al.,}{{Fallows}
  et~al.}{2020a}]{2020JSWSC..10...10F}
{Fallows} R.~A.,  et~al., 2020a, \mn@doi [Journal of Space Weather and Space
  Climate] {10.1051/swsc/2020010}, \href
  {https://ui.adsabs.harvard.edu/abs/2020JSWSC..10...10F} {10, 10}

\bibitem[\protect\citeauthoryear{{Fallows} et~al.,}{{Fallows}
  et~al.}{2020b}]{2020arXiv200304013F}
{Fallows} R.~A.,  et~al., 2020b, J. Space Weather Space Clim., \href
  {https://ui.adsabs.harvard.edu/abs/2020arXiv200304013F} {10,
  arXiv:2003.04013}

\bibitem[\protect\citeauthoryear{{Garcia}, {Kelley}, {Makela}  \&
  {Huang}}{{Garcia} et~al.}{2000}]{2000JGR...10518407G}
{Garcia} F.~J.,  {Kelley} M.~C.,  {Makela} J.~J.,   {Huang} C.~S.,  2000,
  \mn@doi [\jgr] {10.1029/1999JA000305}, \href
  {https://ui.adsabs.harvard.edu/abs/2000JGR...10518407G} {105, 18,407}

\bibitem[\protect\citeauthoryear{{Helfand}, {White}  \& {Becker}}{{Helfand}
  et~al.}{2015}]{2015ApJ...801...26H}
{Helfand} D.~J.,  {White} R.~L.,   {Becker} R.~H.,  2015, \mn@doi [\apj]
  {10.1088/0004-637X/801/1/26}, \href
  {https://ui.adsabs.harvard.edu/abs/2015ApJ...801...26H} {801, 26}

\bibitem[\protect\citeauthoryear{{Hill} \& {Frehlich}}{{Hill} \&
  {Frehlich}}{1996}]{1996ApOpt..35..986H}
{Hill} R.~J.,  {Frehlich} R.~G.,  1996, \mn@doi [\ao] {10.1364/AO.35.000986},
  \href {https://ui.adsabs.harvard.edu/abs/1996ApOpt..35..986H} {35, 986}

\bibitem[\protect\citeauthoryear{{Hocke} \& {Schlegel}}{{Hocke} \&
  {Schlegel}}{1996}]{1996AnGeo..14..917H}
{Hocke} K.,  {Schlegel} K.,  1996, \mn@doi [Annales Geophysicae]
  {10.1007/s00585-996-0917-6}, \href
  {https://ui.adsabs.harvard.edu/abs/1996AnGeo..14..917H} {14, 917}

\bibitem[\protect\citeauthoryear{{Huang}, {Andre}  \& {Sofko}}{{Huang}
  et~al.}{1998}]{1998JGR...103.2131H}
{Huang} C.-S.,  {Andre} D.~A.,   {Sofko} G.~J.,  1998, \mn@doi [\jgr]
  {10.1029/97JA02755}, \href
  {https://ui.adsabs.harvard.edu/abs/1998JGR...103.2131H} {103, 2131}

\bibitem[\protect\citeauthoryear{Jones, Oliphant, Peterson  et~al.}{Jones
  et~al.}{01  }]{citescipy}
Jones E.,  Oliphant T.,  Peterson P.,   et~al., 2001--, {SciPy}: Open source
  scientific tools for {Python}, \url {http://www.scipy.org/}

\bibitem[\protect\citeauthoryear{Kuiack, Huizinga, Molenaar, Prasad, Rowlinson
  \& Wijers}{Kuiack et~al.}{2018}]{10.1093/mnras/sty2810}
Kuiack M.,  Huizinga F.,  Molenaar G.,  Prasad P.,  Rowlinson A.,   Wijers R.
  A. M.~J.,  2018, \mn@doi [Monthly Notices of the Royal Astronomical Society]
  {10.1093/mnras/sty2810}, 482, 2502

\bibitem[\protect\citeauthoryear{{Kuiack}, {Wijers}, {Rowlinson}, {Shulevski},
  {Huizinga}, {Molenaar}  \& {Prasad}}{{Kuiack}
  et~al.}{2020a}]{2020arXiv200300720K}
{Kuiack} M.~J.,  {Wijers} R. A.~M.~J.,  {Rowlinson} A.,  {Shulevski} A.,
  {Huizinga} F.,  {Molenaar} G.,   {Prasad} P.,  2020a, arXiv e-prints, \href
  {https://ui.adsabs.harvard.edu/abs/2020arXiv200300720K} {p. arXiv:2003.00720}

\bibitem[\protect\citeauthoryear{{Kuiack}, {Wijers}, {Shulevski}, {Rowlinson},
  {Huizinga}, {Molenaar}  \& {Prasad}}{{Kuiack}
  et~al.}{2020b}]{2020arXiv200313289K}
{Kuiack} M.,  {Wijers} R. A.~M.~J.,  {Shulevski} A.,  {Rowlinson} A.,
  {Huizinga} F.,  {Molenaar} G.,   {Prasad} P.,  2020b, arXiv e-prints, \href
  {https://ui.adsabs.harvard.edu/abs/2020arXiv200313289K} {p. arXiv:2003.13289}

\bibitem[\protect\citeauthoryear{{Lane}, {Cotton}, {van Velzen}, {Clarke},
  {Kassim}, {Helmboldt}, {Lazio}  \& {Cohen}}{{Lane}
  et~al.}{2014}]{2014MNRAS.440..327L}
{Lane} W.~M.,  {Cotton} W.~D.,  {van Velzen} S.,  {Clarke} T.~E.,  {Kassim}
  N.~E.,  {Helmboldt} J.~F.,  {Lazio} T.~J.~W.,   {Cohen} A.~S.,  2014, \mn@doi
  [\mnras] {10.1093/mnras/stu256}, \href
  {http://adsabs.harvard.edu/abs/2014MNRAS.440..327L} {440, 327}

\bibitem[\protect\citeauthoryear{{Lazio} et~al.,}{{Lazio}
  et~al.}{2010}]{2010AJ....140.1995L}
{Lazio} T. J.~W.,  et~al., 2010, \mn@doi [\aj] {10.1088/0004-6256/140/6/1995},
  \href {https://ui.adsabs.harvard.edu/abs/2010AJ....140.1995L} {140, 1995}

\bibitem[\protect\citeauthoryear{{Loi} et~al.,}{{Loi}
  et~al.}{2015}]{2015GeoRL..42.3707L}
{Loi} S.~T.,  et~al., 2015, \mn@doi [Geophysical Research Letters]
  {10.1002/2015GL063699}, \href
  {https://ui.adsabs.harvard.edu/#abs/2015GeoRL..42.3707L} {42, 3707}

\bibitem[\protect\citeauthoryear{{Loi} et~al.,}{{Loi}
  et~al.}{2016}]{2016JGRA..121.1569L}
{Loi} S.~T.,  et~al., 2016, \mn@doi [Journal of Geophysical Research (Space
  Physics)] {10.1002/2015JA022052}, \href
  {https://ui.adsabs.harvard.edu/abs/2016JGRA..121.1569L} {121, 1569}

\bibitem[\protect\citeauthoryear{McKinney}{McKinney}{2010}]{mckinney-proc-scipy-2010}
McKinney W.,  2010, in van~der Walt S.,  Millman J.,  eds, Proceedings of the
  9th Python in Science Conference. pp 51 -- 56

\bibitem[\protect\citeauthoryear{Molenaar \& Smirnov}{Molenaar \&
  Smirnov}{2018}]{molenaar2018kern}
Molenaar G.,  Smirnov O.,  2018, Astronomy and Computing

\bibitem[\protect\citeauthoryear{{Obenberger} et~al.,}{{Obenberger}
  et~al.}{2015}]{2015JAI.....450004O}
{Obenberger} K.~S.,  et~al., 2015, \mn@doi [Journal of Astronomical
  Instrumentation] {10.1142/S225117171550004X}, \href
  {http://adsabs.harvard.edu/abs/2015JAI.....450004O} {4, 1550004}

\bibitem[\protect\citeauthoryear{{Prasad}, {Wijnholds}, {Huizinga}  \&
  {Wijers}}{{Prasad} et~al.}{2014}]{2014A&A...568A..48P}
{Prasad} P.,  {Wijnholds} S.~J.,  {Huizinga} F.,   {Wijers} R.~A.~M.~J.,  2014,
  \mn@doi [\aap] {10.1051/0004-6361/201423668}, \href
  {http://adsabs.harvard.edu/abs/2014A%26A...568A..48P} {568, A48}

\bibitem[\protect\citeauthoryear{{Prasad} et~al.,}{{Prasad}
  et~al.}{2016}]{2016JAI.....541008P}
{Prasad} P.,  et~al., 2016, \mn@doi [Journal of Astronomical Instrumentation]
  {10.1142/S2251171716410087}, \href
  {http://adsabs.harvard.edu/abs/2016JAI.....541008P} {5, 1641008}

\bibitem[\protect\citeauthoryear{{Price-Whelan} et~al.,}{{Price-Whelan}
  et~al.}{2018}]{astropy:2018}
{Price-Whelan} A.~M.,  et~al., 2018, \mn@doi [\aj] {10.3847/1538-3881/aabc4f},
  \href {https://ui.adsabs.harvard.edu/#abs/2018AJ....156..123T} {156, 123}

\bibitem[\protect\citeauthoryear{{Rowlinson} et~al.,}{{Rowlinson}
  et~al.}{2019}]{2019A&C....27..111R}
{Rowlinson} A.,  et~al., 2019, \mn@doi [Astronomy and Computing]
  {10.1016/j.ascom.2019.03.003}, \href
  {https://ui.adsabs.harvard.edu/abs/2019A&C....27..111R} {27, 111}

\bibitem[\protect\citeauthoryear{{Stewart et al.}}{{Stewart et
  al.}}{2016}]{2016MNRAS.456.2321S}
{Stewart et al.} 2016, \mn@doi [\mnras] {10.1093/mnras/stv2797}, \href
  {http://adsabs.harvard.edu/abs/2016MNRAS.456.2321S} {456, 2321}

\bibitem[\protect\citeauthoryear{{Swinbank} et~al.,}{{Swinbank}
  et~al.}{2015}]{2015A&C....11...25S}
{Swinbank} J.~D.,  et~al., 2015, \mn@doi [Astronomy and Computing]
  {10.1016/j.ascom.2015.03.002}, \href
  {http://adsabs.harvard.edu/abs/2015A%26C....11...25S} {11, 25}

\bibitem[\protect\citeauthoryear{{Thome}}{{Thome}}{1968}]{1968JGR....73.6319T}
{Thome} G.,  1968, \mn@doi [\jgr] {10.1029/JA073i019p06319}, \href
  {https://ui.adsabs.harvard.edu/abs/1968JGR....73.6319T} {73, 6319}

\bibitem[\protect\citeauthoryear{{Tu} \& {Song}}{{Tu} \&
  {Song}}{2016}]{2016JGRA..12111861T}
{Tu} J.,  {Song} P.,  2016, \mn@doi [Journal of Geophysical Research (Space
  Physics)] {10.1002/2016JA023393}, \href
  {https://ui.adsabs.harvard.edu/abs/2016JGRA..12111861T} {121, 11,861}

\bibitem[\protect\citeauthoryear{{Varghese}, {Obenberger}, {Dowell}  \&
  {Taylor}}{{Varghese} et~al.}{2019}]{2019ApJ...874..151V}
{Varghese} S.~S.,  {Obenberger} K.~S.,  {Dowell} J.,   {Taylor} G.~B.,  2019,
  \mn@doi [\apj] {10.3847/1538-4357/ab07c6}, \href
  {https://ui.adsabs.harvard.edu/abs/2019ApJ...874..151V} {874, 151}

\bibitem[\protect\citeauthoryear{{Wijnholds} \& {van der Veen}}{{Wijnholds} \&
  {van der Veen}}{2009}]{2009ITSP...57.3512W}
{Wijnholds} S.~J.,  {van der Veen} A.-J.,  2009, \mn@doi [IEEE Transactions on
  Signal Processing] {10.1109/TSP.2009.2022894}, \href
  {https://ui.adsabs.harvard.edu/abs/2009ITSP...57.3512W} {57, 3512}

\bibitem[\protect\citeauthoryear{{Yao}, {Manchester}  \& {Wang}}{{Yao}
  et~al.}{2017}]{2017ApJ...835...29Y}
{Yao} J.~M.,  {Manchester} R.~N.,   {Wang} N.,  2017, \mn@doi [\apj]
  {10.3847/1538-4357/835/1/29}, \href
  {http://adsabs.harvard.edu/abs/2017ApJ...835...29Y} {835, 29}

\bibitem[\protect\citeauthoryear{{de Gasperin}, {Mevius}, {Rafferty}, {Intema}
  \& {Fallows}}{{de Gasperin} et~al.}{2018}]{2018A&A...615A.179D}
{de Gasperin} F.,  {Mevius} M.,  {Rafferty} D.~A.,  {Intema} H.~T.,   {Fallows}
  R.~A.,  2018, \mn@doi [\aap] {10.1051/0004-6361/201833012}, \href
  {https://ui.adsabs.harvard.edu/abs/2018A&A...615A.179D} {615, A179}

\bibitem[\protect\citeauthoryear{{van Haarlem} et~al.,}{{van Haarlem}
  et~al.}{2013}]{2013A&A...556A...2V}
{van Haarlem} M.~P.,  et~al., 2013, \mn@doi [\aap]
  {10.1051/0004-6361/201220873}, 556, A2

\bibitem[\protect\citeauthoryear{{van der Walt}, {Colbert}  \&
  {Varoquaux}}{{van der Walt} et~al.}{2011}]{2011CSE....13b..22V}
{van der Walt} S.,  {Colbert} S.~C.,   {Varoquaux} G.,  2011, \mn@doi
  [Computing in Science and Engineering] {10.1109/MCSE.2011.37}, \href
  {https://ui.adsabs.harvard.edu/abs/2011CSE....13b..22V} {13, 22}

\makeatother
\end{thebibliography}








\bsp	
\label{lastpage}
\end{document}